\documentclass{eas}
\usepackage{graphicx}
\usepackage{url}

\newcommand{\apj}{ApJ}
\newcommand{\mnras}{MNRAS}
\newcommand{\aap}{A\&A}

\begin{document}

\title{Radiative transfer in protoplanetary disks} 
\author{C. Pinte}\address{Laboratoire d'Astrophysique de Grenoble, CNRS/UJF UMR~5571, 
414 rue de la Piscine, B.P. 53, F-38041 Grenoble Cedex 9, France}
\author{F. M\'enard$^1$}
\author{G. Duch\^ene$^1$}
\begin{abstract}
We present a new 3D continuum radiative transfer  code, 
MCFOST, based on a Monte-Carlo method. The reliability and efficiency
of the code is tested by comparison with five different radiative transfer codes
previously tested by \cite{Pascucci04}, using a 2D disk
configuration.  When tested against the same disk configuration,
no significant difference is found between the temperature and SED
calculated with MCFOST and with the other codes.  The
computed values are well within the range of values computed by the
other codes. The code-to-code differences are small, they rarely
exceed 10\% and are usually much smaller. 
\end{abstract}
\maketitle

\section{Introduction}
Dust is present everywhere in astrophysics, from the
interstellar medium to the atmospheres and close circumstellar
environments of numerous classes of objects; from the lower mass
planets and brown dwarfs, to massive stars. With the advent of high
angular resolution and high contrast imagers, the basic structural
properties (e.g., size, inclination, and surface brightness) of the
circumstellar environments of the nearest and/or largest objects ---
disks and envelopes around young stars in nearby star forming
regions and around more distant evolved stars --- are now under close
scrutiny. 
With this unprecedented wealth of high resolution data,
from optical to radio, fine studies of the dust content become
possible and potent radiative transfer (RT) codes are needed to fully
exploit the data.

At short wavelengths, dust grains efficiently absorb, scatter, and
polarise the starlight while at longer wavelengths dust re-emits the
absorbed radiation. How much radiation is scattered and absorbed is a
function of both the geometry of the circumstellar environment and the
properties of the dust. In turns, the amount of absorbed radiation
sets the temperature of the dust (and gas) and defines the amount of
radiation that is re-emitted at longer, thermal, wavelengths.

To get a reliable understanding of the structure and evolution of
these ``dusty'' objects, be it the evolution of dust sizes, the
temperature dependent chemistry, or simply the density profiles, it is
highly desirable to model not only the integrated fluxes (\textit{i.e.}, the
spectral energy distributions), but also the resolved brightness maps
and polarisation profiles when available. This can only be done by
solving the radiative transfer (hereafter RT) problem in media that
can have large optical depths and/or complex geometries and
compositions. Needless to say, analytical solutions do not always
exist and sophisticated numerical methods must generally be used. One
such versatile numerical method is the Monte Carlo method. The code we
describe and test below is based on that RT scheme.

The original version of the code, MCFOST, treated the polarised
continuum RT problem by including scattering only, without a
calculation of the temperature structure and thermal emission from the
circumstellar environment \cite{Menard}. That code was used
extensively to produce synthetic images of the scattered light from
disks around young stars. Examples include the circumbinary ring of
GG~Tau \cite{Duchene04}, the large silhouette disk associated with
IRAS 04158+2805 (M\'enard et al. 2005), and an analysis of the circular
polarisation in GSS~30 \cite{Chrysostomo97}. In this paper we present
a new improved version of MCFOST that includes passive heating of the
circumstellar environment by stellar radiation and thermal re-emission
from the dust, needed to prepare future observations with Spitzer,
Hershel and Alma.

To validate MCFOST, we perform a comparison with five other RT codes,
including three using a Monte Carlo method that were benchmarked in a
previous study \cite{Pascucci04} (hereafter P04). To compare the
reliability and efficiency of each code, P04 used a well defined
2-dimensional disk configuration, with simple dust properties. Each
code calculated the temperature structure in the disk, as well as the
emergent SED. The results were compared quantitatively as a function
of optical thickness and disk inclination. The agreement is
generally better than 10\% for the most difficult cases, \textit{i.e.}, large
optical depths, and much better in the optically thin cases. We
performed identical calculations with MCFOST and present the results
and comparisons below.

In \S2, we briefly describe MCFOST. In section \S3, we recall the
benchmark problem used to compare MCFOST with the 5 RT codes included
in P04. Results are presented in \S4, they include the temperature
profile and SED for a range of disk optical thicknesses.

\section{Description of the numerical code}
MCFOST is a 3D continuum radiative transfer numerical code based on
the Monte-Carlo method.  It was originally developed by \cite{Menard}
to model the scattered light in dusty media (including linear and
circular polarisations). In this paper, we present an extended version
of the code that includes dust heating and continuum thermal
re-emission.

\subsection{Geometry of the computation grid}

The spatial grid is defined in cylindrical
coordinates. It is well adapted to the geometry of circumstellar
disks. We use $N_r$ logarithmically spaced radial grid cells and $N_z$
linearly spaced vertical grid cells. The size of the vertical cells
follow the flaring of the disk, with a cutoff at 10~times the local
scale height. For the purpose of the benchmark calculations presented
in this paper, axisymmetry is assumed. However, MCFOST allows the
density distribution to be defined arbitrarily in 3D, with the
limitation that within each cell, quantities are held constant. For
example, MCFOST
in a 3D version has already been used to the study the
photo-polarimetric behaviour of AA~Tau \cite{Pinte04}.

\subsection{Position dependence of the dust distribution}
An explicit spatial dependence of the size distribution $f(a,\vec{r})$
is implemented in MCFOST.  The dust properties are therefore defined
locally, \textit{i.e.}, each cell of the disk may contain its own independent dust
population. This allows to model dust settling towards the
midplane of the disk and/or variations of the chemical composition of
the dust from the inner, hot regions to the outer, cold edge
of the disk.

\subsection{Optical dust properties}
Dust particles can be arbitrary shaped and are assumed to be randomly
oriented. In the case of homogeneous and spherical particles,  
the dust optical properties are computed with Mie theory. 
The optical
properties in any cell of the disk are derived in accordance with the
local size and composition distributions $f(a,\vec{r})$, at the
expense of a significant usage of memory space. The extinction and
scattering opacities are given by
\begin{equation} 
\kappa^\mathrm{ext/sca}(\lambda,\vec{r}) =
\int_{a_\mathrm{min}}^{a_\mathrm{max}} \pi a^2
Q_\mathrm{ext/sca}(\lambda,a) f(a,\vec{r}) \mathrm{d}a 
\end{equation}
and the local Muller matrix is 
\begin{equation}
\label{eq:def_S}
S(\lambda,\vec{r}) = \int_{a_\mathrm{min}}^{a_\mathrm{max}}  S(\lambda,a) f(a,\vec{r}) \mathrm{d}a
\end{equation}
where $ Q_\mathrm{sca}(\lambda,a)$, $ Q_\mathrm{ext}(\lambda,a)$ and
$S(\lambda,a)$ are respectively the scattering and extinction cross sections and
the Mueller matrix of a grain of size $a$ at a wavelength $\lambda$.

\subsection{The radiative transfer scheme in MCFOST}

The Monte Carlo method allows to follow individual
monochromatic ``photon packets'' that propagate through the
circumstellar environment until they exit the computation grid.  The
propagation process is governed by scattering, absorption and
re-emission events that are controlled by the optical properties of
the medium (opacity, albedo, scattering phase function, etc...) and by
the temperature distribution. Upon leaving the circumstellar
environment, photon packets are recorded into frequency dependent
synthetic images and SED's can be constructed.

To improve its computational efficiency, MCFOST uses different
strategies to set the energy of a photon packet, corresponding to
different samplings of the radiation field. The energy of a
packet, hence the number of photons it contains, is chosen and
optimised depending on the goal of the calculations. On the one hand,
the convergence of the temperature distribution is optimized when all
photon packets have the same {\sl energy}, independently of their
wavelengths. This procedure insures that more photon packets are
emitted in the more luminous bins. On the other hand, the computation
of SEDs is more efficient when it is the {\sl number} of photon
packets that is held constant for all wavelength bins. In that case
it is the energy of the packets that is wavelength dependent.
Thus, MCFOST is more efficient when it computes the temperature and
SED with a two-step process:
\begin{itemize}
\item {\bf Step 1} is the temperature determination. Photon packets
are generated and calculated one by one. They are generated at the
stellar photosphere and followed until they exit the computation
grid.  Upon scattering in the disk, the propagation vector of the
packet is modified, but not its wavelength. Upon absorption however,
packets are immediately re-emitted, in situ and isotropically, but at a
different wavelength, calculated according to the temperature of the
grid cell. For this re-emission process, the concept of {\sl
immediate reemission} and the associated {\sl temperature correction
method} proposed by \cite{Bjorkman01} are used. They are presented as
\emph{scheme 3} in P04. In this step, all photon packets have a
similar energy and are randomly scattered/absorbed within the
disk. The concept of {\sl mean intensity} suggested by \cite{Lucy99}
is further used to reduce noise in the temperature estimation for
optically thin cases. Step 1 allows for a fast convergence of the
temperature but is rather time-consuming when used to derive SEDs,
especially in the low-energy, long wavelength regime. Therefore SEDs are
calculated differently.

\item {\bf Step 2} computes the SED and/or images from the 
temperature distribution calculated in step 1.  In step 2, the number
of photon packets is held constant at all wavelengths and the {\sl
enforced scattering} concept of \cite{Cashwell} is implemented to
further speed up convergence. Step 2 maintains a similar noise level
per wavelength bin and reduces convergence time for the SED by
limiting the CPU time spent in high luminosity bins and focussing on
low luminosity bins.
\end{itemize}

\subsubsection{The emission of photon packets}

In MCFOST, two radiation sources are considered, the star and the
circumstellar environment. The stellar photosphere can be represented
by (i) a sphere radiating uniformly or (ii) a limb-darkened sphere or
(iii) a point-like source, whichever is relevant for the problem under
consideration.  Hot and/or cold spots can be added the photosphere. In
the following, we consider a unique central point-like star that
radiates like a black body. 

In step 1 of the previous subsection, all photon packets are
emitted isotropically by the central star and have the same energy,
independently of their wavelength. We define the luminosity of a
packet, $\epsilon$, as:
\begin{equation} 
\epsilon = L_*/N_{\gamma step1} 
\end{equation} 
where $L_*$ is the bolometric luminosity of the star and $ N_{\gamma
step1}$ the number of packets generated. The wavelength of the packet
is chosen according to a normalised probability density function
proportional to $B_\lambda(T_*)$.

In step 2, the SED is constructed using a constant number $N_{\gamma
step2}$ of packets for all the wavelength bins. The packets are
emitted by the star and by the disk. The energy of the $N_{\gamma
step2}$ packets emitted at a given wavelength $\lambda$ is determined
by the total energy that the star and the disk radiates at this
wavelength. The corresponding packet luminosity is~:
\begin{equation}
\epsilon_\lambda = \frac{L_*(\lambda)~+~\sum_i~L_i(\lambda)}
{N_{\gamma step2}}
\end{equation}
where $i$ is the index of the cell and $L_*(\lambda)$ and
$L_i(\lambda)$ are the luminosities of the star and disk cell at a
given wavelength. They are given by:

\begin{equation}
L_*(\lambda) = 4 \pi^2R_*^2\,B_\lambda(T_*)~,
\end{equation}

\begin{equation}
{\rm and}~ L_i(\lambda) = 4 \pi \, \kappa_i^{\mathrm{abs}}(\lambda) \,B_\lambda(T_i) m_i~.
\end{equation}

The wavelength is set in a deterministic way and packets are randomly
emitted from the star and the disk, by cell $i$, with the respective
probabilities:
\begin{equation} 
p_* = \frac{L_*(\lambda)}{L_*(\lambda)~+~\sum_i~L_i(\lambda)}~,
\end{equation}

\begin{equation}
{\rm and}~ p_i = \frac{L_i(\lambda)}{L_*(\lambda)~+~\sum_i~L_i(\lambda)}~.
\end{equation}

In each disk cell we assume that the density, temperature and
opacities are constant. To determine the position of emission of a photon
within a given disk cell, in cylindrical geometry, the following relations
are used:
\begin{equation}
r = \sqrt{r_\mathrm{min}^2 + A (r_\mathrm{max}^2 -  r_\mathrm{min}^2)}~,
\end{equation}
\begin{equation}
|z| = z_\mathrm{min} + A (z_\mathrm{max} -  z_\mathrm{min})~,
\end{equation}
\begin{equation}
{\rm and}~ \phi = 2 \pi A
\end{equation}
where A denotes three different random numbers drawn from a uniform
distribution in the interval ]0,1] and where $r_\mathrm{min}$,
$r_\mathrm{max}$, $z_\mathrm{min}$, $z_\mathrm{max}$ are the radial
and vertical boundaries of the cell.  The dust thermal emission is assumed
isotropic.

\subsubsection{Distance between interactions}

It is natural within a Monte Carlo scheme to estimate the distance a
photon packet ``travels'' between two interactions by means of optical
depth (related naturally to the density) rather than by physical, linear
distance. From a site of interaction, the optical depth
to the next site is randomly chosen from:
\begin{equation}
\tau_\lambda = -\ln A
\end{equation}
with $A \in ~ ]0,1]$.
The distance $l$ is computed by integrating the infinitesimal
optical depth $\kappa^\mathrm{ext}(\lambda,\vec{r}) \rho(\vec{r})
\mathrm{d}s$ until the following equality is verified~:
\begin{equation}
\tau_\lambda = \int_0^l \kappa^\mathrm{ext}(\lambda,\vec{r}) \rho(\vec{r}) \mathrm{d}s~.
\end{equation}

Once the position of interaction $\vec{r}$ is determined, the
probability that the interaction is a scattering event rather than an 
absorption event is estimated with the local (\textit{i.e.}, cell) albedo~:
\begin{equation}
{p_\mathrm{sca} = \mathcal{A}(\lambda,\vec{r}) = \frac{\int \pi a^2
Q_\mathrm{sca}(\lambda,a) f(a,\vec{r}) \mathrm{d}a}{\int \pi a^2
Q_\mathrm{ext}(\lambda,a) f(a,\vec{r}) \mathrm{d}a}}~.
\end{equation}
In step 1, the photon packets are scattered or absorbed following this
probability whereas in step 2, they are always scattered, no
absorption is allowed to occur explicitely. Instead, scattering is
enforced at each interaction site but the Stokes vector is weighted
 by the probability of scattering, $p_\mathrm{sca}$. This
allows all the photons to exit the disk and to contribute to the SED,
although with a reduced weight.

\subsubsection{Scattering}

MCFOST includes a complete treatment of polarisation, including linear
and circular polarisations. In the Stokes formalism used, the state of
a light packet is described by its Stokes
vector, with $I$ the intensity, $Q$ and $U$ describing the linear
polarisation, and $V$ the circular polarisation.  The interaction of a
photon with a dust particle is described by a
$4\times4$ matrix, the Mueller matrix, $S$. In the simplifying case
where dust particles are homogeneous and spherical (Mie Theory), the
matrix becomes block-diagonal, only 4 different elements are non-zero,
and only three of these are independent.

\begin{equation}
\label{eq:mueller}
\left( \begin{array}{r}
I \\
Q \\
U \\
V \\
\end{array} \right)_{i+1}
=
\left( \begin{array}{rrrr}
S_{11} & S_{12} & 0 & 0 \\
S_{12} & S_{11} & 0 & 0 \\
0    &  0  & S_{33} & S_{34} \\
0 & 0 & -S_{34} & S_{33} \\
\end{array} \right)
\left( \begin{array}{r}
I \\
Q \\
U \\
V \\
\end{array} \right)_{i}
\end{equation}
The direction of scattering is defined by two angles $\theta$ and
$\phi$ in spherical coordinates. Each individual element $S_{ij}$ of
the matrix is a function of the two scattering angles and of the
wavelength, $S_{ij} = S_{ij}(\theta,\phi,\lambda)$

The scattering angle $\theta$ is
randomly chosen from the pretabulated scattering phase function~:
\begin{equation}
\label{eq:phase_theta}
\Phi_\lambda(\theta) \propto S_{11} (\theta)
\end{equation} 
where $S_{11}$ is the first element of the Mueller matrix.

For unpolarised incoming light packet, \textit{i.e.}, $Q,U = 0$, the
distribution of azimuthal angle $\phi$ is isotropic. For light packets
with a non-zero linear polarisation $P =\sqrt{Q^2+U^2}/I$, the
azimuthal angle is defined relative to its direction of polarisation
and determined by means of the following repartition function~:
\begin{equation}
\label{eq:phase_phi}
F_\theta(\phi) =
\frac{1}{2\pi}\left(\phi-\frac{S_{11}(\theta)-S_{12}(\theta)}{S_{11}(\theta)+S_{12}(\theta)}\,P\,\frac{\sin(2\phi)}{2}\right)
\end{equation} 
where $\theta$ was previously chosen from equation
\ref{eq:phase_theta} (see \cite{Solc89})

The Mueller matrix used in equation \ref{eq:mueller},
\ref{eq:phase_phi} and  \ref{eq:phase_theta} can be defined in two different 
ways : one Mueller matri xper grain size or one mean Mueller matrix per
grid cell. In the latter case, the Mueller matrix is defined by
equation \ref{eq:def_S}. In the former case, the grain size must be
chosen explicitely for each event following the probability law~:
\begin{equation}
p(a)da = \frac{\pi a^2 Q_{\mathrm{sca}}(a)f(a,\vec{r})da}{\int_{a_\mathrm{min}}^{a_\mathrm{max}}\pi a^2 Q_{\mathrm{sca}}(a)f(a,\vec{r})da}~.
\end{equation}
The two method are strictly equivalent. They can be used alternatively
to optimise either the memory space or the computation time required.

\subsubsection{Absorption and radiative equilibrium\label{sec:eq_rad}}
The temperature of the dust particles is determined by assuming radiative equilibrium and that the dust opacities do not depend on
temperature.

Each time a packet $\gamma$ of a given wavelength $\lambda$ travels
through a cell, we compute $\Delta l_\gamma$, the length
travelled by the packet in the cell and the mean intensity
$J_\lambda$ in the cell is derived following \cite{Lucy99} :

\begin{equation}
J_\lambda = \frac{1}{4\pi V_i} \sum_{\gamma} \epsilon\, \Delta
l_\gamma = \frac{1}{4\pi V_i} \sum_{\gamma} \frac{L_*}{N_\gamma}\,
\Delta l_\gamma
\end{equation}
where $V_i$ is the volume of the cell $i$.

The dust thermal balance should take into account
the thermal coupling between gas and dust. In high density regions,
close to the disk midplane, the coupling is very strong and the gas
temperature should be close to the dust temperature. In the surface
layers of the disk, on the other hand, density becomes very low and
gas-dust thermal exchanges should be reduced. 

The two extreme assumptions are implemented in MCFOST~: either the
gas-dust thermal exchange is perfectly efficient and gas and dust are in local
thermodynamic equilibrium (LTE), either there is no thermal coupling
between gas and dust.

If we assume local thermodynamic equilibrium, all dust grains
have the same temperature which is equal to the gas temperature. We
can define this temperature as the temperature of
the cell, given by the radiative equilibrium equation~:
\begin{equation}
4\pi\int_0^\infty \kappa_i^\mathrm{abs}(\lambda) B_\lambda(T_i)
d\lambda = 4\pi\int_0^\infty \kappa_i^\mathrm{abs}(\lambda)
J_\lambda d\lambda
\end{equation}
which we write~: 
\begin{equation} 
\int_0^\infty \kappa_i^\mathrm{abs}(\lambda)
B_\lambda(T_i) d\lambda = \frac{L_*}{4\pi V_i N_\gamma} \sum_{\lambda,
\gamma} \kappa_i^\mathrm{abs}(\lambda) \Delta l_\gamma~. 
\end{equation}

If there is no thermal coupling (other than radiative) between gas and
dust, and then between the
different dust grain sizes, each grain size has its
own temperature and the radiative must be, in this case,
written independently for each size~:
\begin{equation}
4\pi\int_0^\infty \kappa_{i}^\mathrm{abs}(\lambda,a) B_\lambda(T_{i}(a))
d\lambda = 4\pi\int_0^\infty \kappa_i^\mathrm{abs}(\lambda,a)
J_\lambda d\lambda
\end{equation}
where $\kappa_{i}^\mathrm{abs}(\lambda,a)$ and $T_{i}(a)$ are the
opacity and temperature of the dust grains of size $a$ in the cell $i$.

The calculation of $\int_0^\infty \kappa_i^\mathrm{abs}(\lambda)
B_\lambda(T_i) d\lambda = \sigma T^4 \kappa_P / \pi$ ($\kappa_P$ is
the Planck mean opacity) is very time consuming and we pretabulate
these values at $N_T = 1000$ logarithmically spaced temperatures
ranging from $1$K to $1500$K. The
temperature $T_i$ of each cell is obtained by interpolation between the
tabulated temperatures.

Following \cite{Bjorkman01}, we use the concept of {\sl immediate
reemission}. When a packet is absorbed, it is immediately reemitted and
its wavelength is chosen taking into account the temperature correction
following the probability distribution :
\begin{equation} p_\lambda d\lambda \propto
\kappa_\mathrm{i}^\mathrm{abs}(\lambda)
\left(\frac{\mathrm{d}B_\lambda}{\mathrm{d}T}\right)_{T_i} d\lambda~.
\end{equation}

\section{\label{sec:bench} Benchmark problem}
In order to test MCFOST, we calculated the same cases as described in
P04 and compared the results. The geometry tested involves a central
point-like source radiating as a T=$5\,800$K blackbody encircled by a
disk of well defined geometry and dust content. The disk extends
from 1 AU to 1000 AU. It includes spherical dust grains made of
astronomical silicate. Grains have a radius of $0.12\,\mu$m and a
density of $3.6\ \mathrm{g.cm}^{-3}$. The optical data are taken from
\cite{Draine84}. The disk is flared with a Gaussian vertical
profile $\rho(r,z) = \rho_0(r)\,exp(-z^2/2\,h(r)^2)$. Power-law
distributions are used for the surface density $\Sigma(r) =
\Sigma_0\,(r/r_0)^{\alpha}$ and the scale height $ h(r) = h_0\,
(r/r_0)^{\beta}$ where $r$ is the radial coordinate in the equatorial
plane, $h_0$ the scale height at the radius $r_0$. These assumptions
lead to a general expression for the density at any point in the disk
: \begin{equation} \rho(r,z) =
\rho_0\,\left(\frac{r}{r_0}\right)^{\alpha-\beta}\,\exp\left(-\frac{1}{2}\left(\frac{z}{h(r)}\right)^2\right)~.
\end{equation}

In this benchmark, $\beta=1.125$, $\alpha-\beta=-1.0$ are used. The
reference radius is $r_0= 500$ AU, the disk height is $h_0 = 99.74$
AU. This scale height is the same as described in P04. A factor
$\sqrt{2/\pi}$ appears because of the different definition of the
Gaussian profile in MCFOST\footnote{See the definition of $f_2(r)$ in
eq.~4 of P04.}.

The scattering is set to be isotropic and polarisation is not
calculated, \textit{i.e}, all the scattering information is contained in the
scattering cross section alone, $Q_{sca}$, acting on the I Stokes
parameter only.
 
Results are shown in the section below for 4 different
disk equatorial optical thicknesses, $\tau_\mathrm{V}$ : 0.1, 1, 10, and 100.
For each optical thickness case, results for three different disk
inclinations  are calculated, 12.5$^\circ$, 42.5$^\circ$ and 77.5$^\circ$.

\section{Results}
\subsection{Computational considerations}
For the test cases considered here, the number of grid cells is set to
$N_r = 50$ and $N_z = 20$ in the radial and vertical directions.  $10^7$ photon packets
are used to calculate the temperature distribution (step 1) and
$2\,10^6$ photon packets per wavelength are used for the generation of
the SED (step 2). The total runtime is $13$ minutes for the most
optically thin case and $20$ minutes for the most optically thick case
on a bi-processor (Intel Xeon) computer running at a clockrate of 2.4
GHz. The runtime memory space needed is $10$ MBytes.

For the high resolution models (see sec. \ref{sec:high_res} below),
with $N_r = 500$ and $N_z = 200$, the runtime for the most optically
thick case is $\approx 7$ hours and the memory occupation is $450$
MBytes.

\subsection{The temperature profiles}
\begin{figure}[!ht]
\centering
\includegraphics[width=8cm,angle=270]{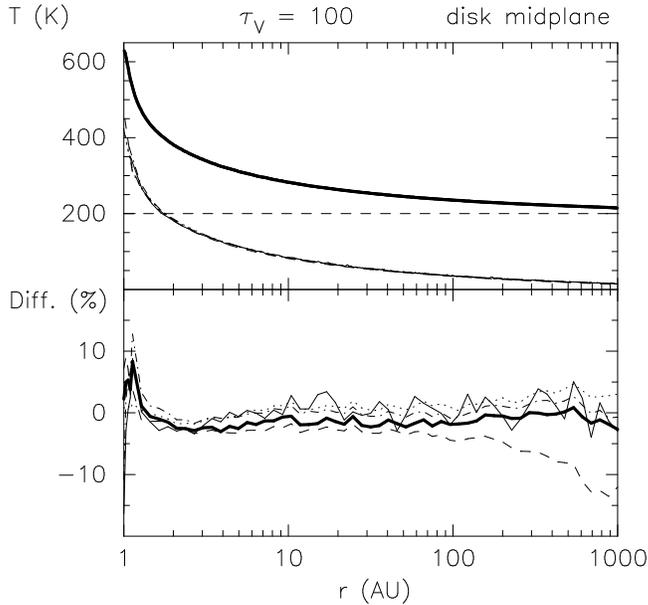}

\caption{Plots of the radial temperature (upper panel) and the 
difference (lower panel) for the most optically thick case, $\tau_\mathrm{V} =
100$ and using RADICAL as the reference. 
In both panels, the results of MCFOST are represented by the thick 
solid line. Thin solid
lines are the results from MC3D, dot-dashed lines from MCTRANSF,
dashed-dot-dot-dot from RADICAL, dotted lines from RADMC and dashed
lines from STEINRAY. In the upper panel, and all curves being very
similar, MCFOST has been shifted by 200K for clarity. 
\label{fig:Temp_radiale}}
\end{figure}

\begin{figure}[!ht]
\centering
\includegraphics[width=8cm,angle=270]{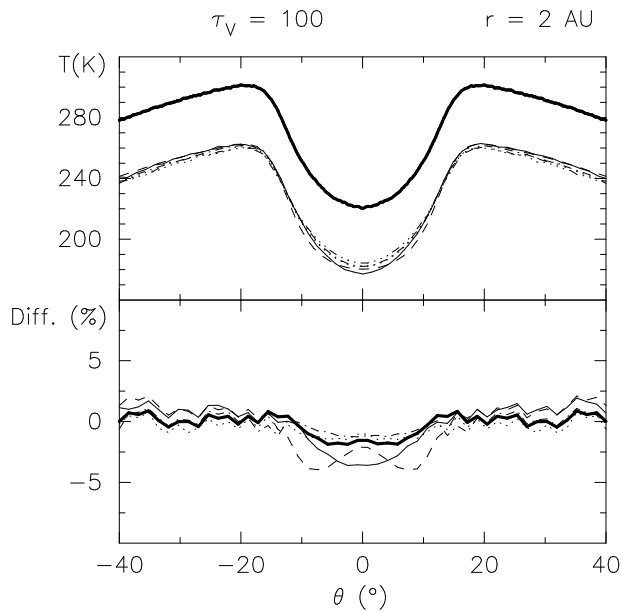}
\caption{Plots of the vertical temperature (upper panel) and the 
difference (lower panel) among the codes using RADICAL as the
reference, for the most optically thick case, $\tau_\mathrm{V} = 100$ and for a
distance $r$ in the midplane equal to 2 AU from the central star. In
the upper panel, the results of MCFOST are shifted by 40 K for clarity. The line
types are the same as in
Fig. \ref{fig:Temp_radiale}.\label{fig:Temp_verticale}}
\end{figure}

The radial and vertical temperature profiles for the most optically
thick case are shown in the top panels of Fig. \ref{fig:Temp_radiale} and
\ref{fig:Temp_verticale}. For clarity the results of MCFOST are
shifted by 200~K in Fig. \ref{fig:Temp_radiale} and by 40~K in Fig.
\ref{fig:Temp_verticale}. The lower panels of both figures show the
difference, given in percents, taking the code RADICAL as a
reference. The thick solid lines traces the difference of MCFOST with
RADICAL. The radial temperature of MCFOST does not differ by more than
5\% from all the other codes, except from MC3D and RADICAL close to
the inner radius of the disk, and from STEINRAY at large radius, but
the maximum difference remains below 15\%. In the vertical direction,
MCFOST always agrees better than 2.5\% with RADICAL, MCTRANSF and
RADMC. The agreement with MC3D and STEINRAY is always better than
2.5\% at high altitudes ($\theta > 20^\circ$). Closer to the midplane
deviations are larger but do not exceed 4\%.

\subsection{The spectral energy distributions}
\begin{figure}[!ht]
\centering
\includegraphics[angle=270,width=9.5cm]{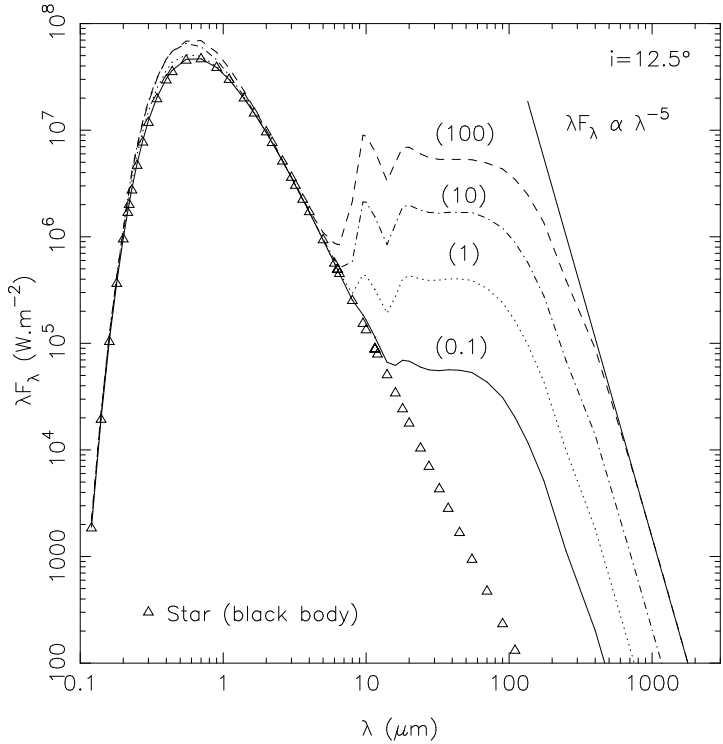}
\caption{\label{fig:sed} SED for a disk inclination $i=12.5^\circ$. Each curve traces a model
SED calculated by MCFOST. The midplane optical depth is given in
parenthesis above each curve. The solid lines show results for the
most optically thin disk, $\tau_{\nu}=0.1$, dotted lines for a disk
having $\tau_{\nu}=1$, dot-dashed lines for a disk with
$\tau_{\nu}=10$, and dashed lines for the most optically thick model,
$\tau_{\nu}=100$. The diamonds trace the black-body emission from the
naked star. The slope of the SED at long wavelengths depends only on
the dust properties and is plotted in each panel with a solid line,
F$_{\lambda} \propto \lambda^{-6}$.}
\end{figure}

Shown in Fig. \ref{fig:sed} are the calculated SEDs for two disk
inclinations and four different optical thickness, $\tau_\mathrm{V}$. The
figure directly compares with Fig.~7 of P04. We plot $\lambda
F_\lambda$ in (W.m$^{-2}$) where $F_\lambda$ is the flux density at a
distance equal to the stellar radius, for a naked star (without
disk) $F_\lambda = \pi B_\lambda$ (triangles in
Fig. \ref{fig:sed}). MCFOST reproduces the correct slope at long
wavelengths, \textit{i.e.}, $\lambda F_\lambda \propto \lambda^{-5}$, expected
for an optically thin medium containing small particles (with $\kappa
\propto \lambda^{-2}$), emitting at long wavelength (B$_{\lambda} 
\propto \lambda^{-4}$).

\begin{figure}[!ht]
\centering
\includegraphics[angle=270,width=9.5cm]{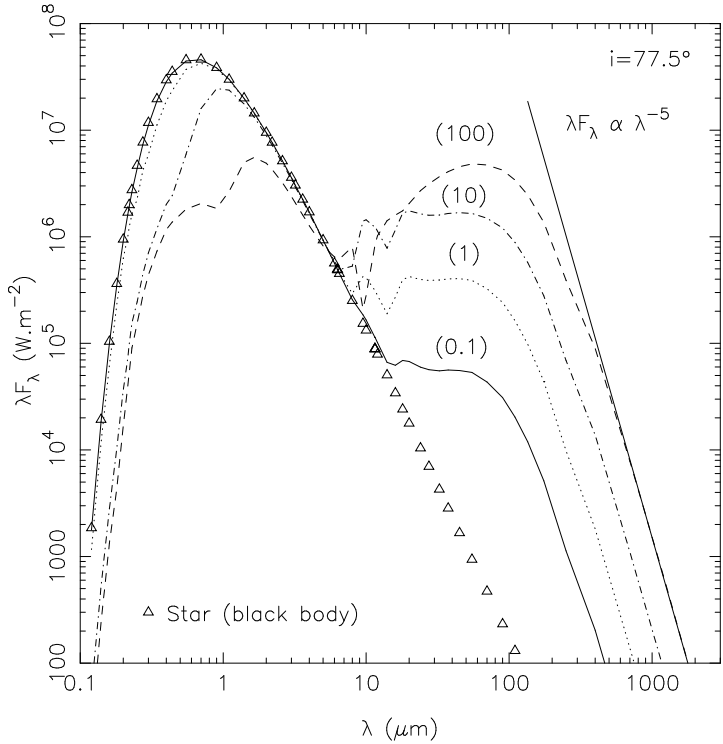}
\caption{\label{fig:sed2} Same as Fig. \ref{fig:sed} for an
  inclination angle of $i=77.5^\circ$}
\end{figure}

\begin{figure}[!ht]
\centering
\includegraphics[angle=270,width=9.5cm]{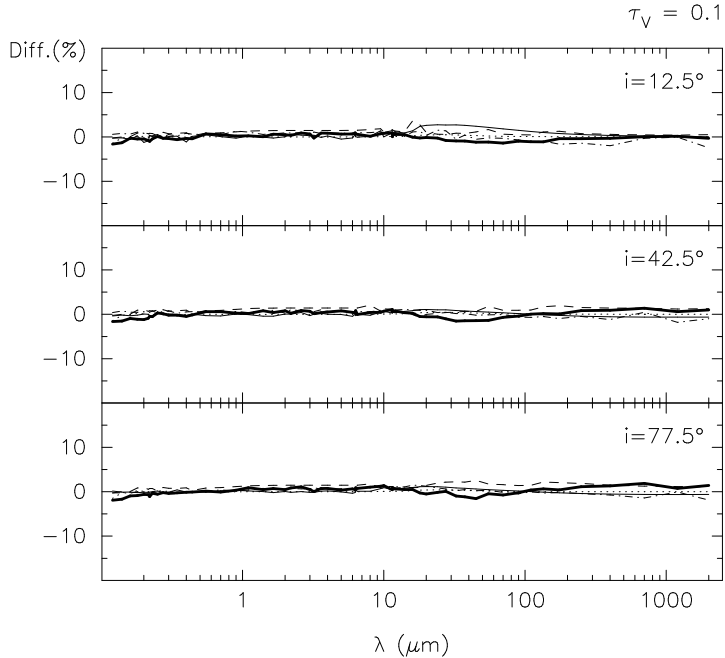}
\caption{\label{fig:diff} Plots of the difference between the model
SED, given in percents, using the code RADICAL as the
reference. Results from all codes are shown: the five codes included
in the study of P04 and MCFOST (this study).  Results are shown as a
function of optical thickness of the disk midplane and as a function
of inclination angle.  Results are shown for the most
optically thin case $\tau_{\nu}=0.1$. 
Three different
inclination angles are considered: $i=12.5^\circ$ (upper), $i=42.5^\circ$
(middle), and $i=77.5^\circ$ (lower). All codes are compared to
RADICAL. The thick solid line gives the differences for MCFOST, the
thin solid line for MC3D, the dot-dashed line for MCTRANSF, the dotted
line for RADMC, and the dashed lines for STEINRAY.}
\end{figure}

In Fig. \ref{fig:diff}, the difference of the SEDs calculated by the
various RT codes are presented. The overall agreement is better at
lower inclinations, a consequence of the lower optical
depth along the observer's line of sight. Indeed, for models with $\tau_\mathrm{V} = 0.1$ and $\tau_\mathrm{V} = 1$ at
all inclinations and for models with $\tau_\mathrm{V} = 10$ and $\tau_\mathrm{V} = 100$
at inclinations of $12.5^\circ$ and $42.5^\circ$, deviations between
MCFOST and all the other codes do not exceed 10\%. For the most
optically thick cases, \textit{i.e.}, large inclination and/or large optical
thickness, the largest differences are observed below $1\mu$m, where
scattering dominates, and around $10
\mu$m, in the silicate band.

In the optical range, MCFOST produces slightly larger fluxes with
respect to other codes but the large dispersion observed in all codes
does reflect lower signal to noise ratio due to lower emerging fluxes.
Near $10 \mu$m, MCFOST shows the same trend as MCTRANSF but finds
larger fluxes than RADICAL or MC3D.

To accelerate convergence in MCFOST, photon packets are recorded in
inclination bins that cover a range of angles, instead of a single
angle as is often done elsewhere. It is important to note that by
doing so, the uncertainty associated with the models easily matches
the observational errors. Observationally, inclination angles of disks are rarely
known to better than 5-10 (1-2) degrees when seen pole-on
(edge-on). In MCFOST, the inclination bins intercept equal solid
angles from pole-on to edge-on, \textit{i.e.}, they cover equal intervals in
$\cos i$. This interval can be set to match the quality of the data.
In the calculations presented in this papers, 21 bins are used from
pole-on to edge-on. In all figures, the inclinations
listed, \textit{i.e.}, $12.5^\circ$, $42.5^\circ$, $77.5^\circ$, are the median
of the bins covering an interval of 0.05 in $\cos i$.  These bins have
inclinations ranging from $0^\circ$ to $17.75^\circ$, from
$40.37^\circ$ to $44.42^\circ$, and from $76.22^\circ$ to
$79.02^\circ$, respectively.

For comparison, the axis ratio of a circular disk estimated with a
measurement error equal to half the bin width used in this paper,
\textit{i.e.}, 0.025, would yield an error on $i$ of $\sim10^\circ$ for the pole-on
case, and $\sim 1.5^\circ$ in the edge-on case. Indeed, for a perfectly
circular pole-on disk, measuring an axis ratio of 0.975 instead of 1.0
would mimic an inclination of $i=12^\circ$ instead of $i=0^\circ$. In the
edge-on case, measuring an axis ratio of 0.1 instead of 0.125 would
induce a difference of 1.4$^\circ$ in the determination of the
inclination. 
The results presented here, generally obtained using 21 inclination
bins, are therefore accurate enough and not limited by our choice of inclination bin number.

\begin{figure}[!ht]
\centering
\includegraphics[angle=270,width=9.5cm]{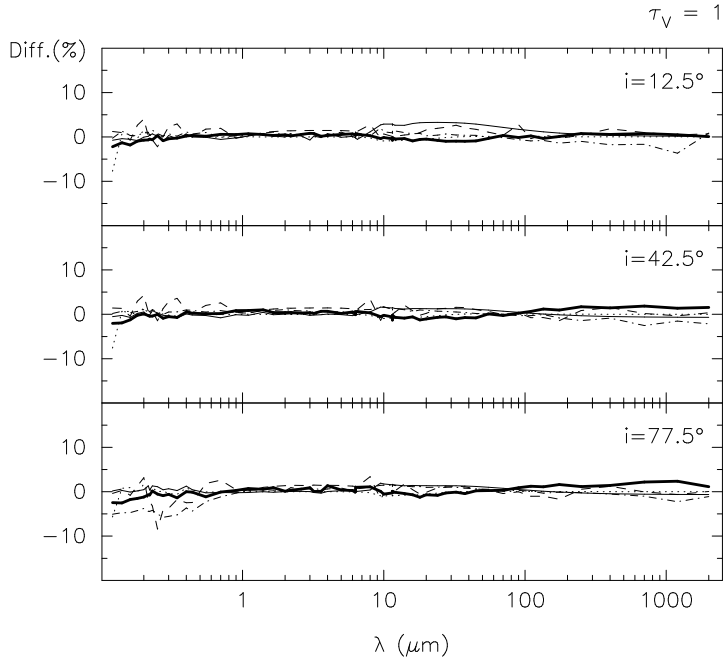}
\caption{\label{fig:diff2} Same as figure \ref{fig:diff} but for $\tau_{\nu}=1$.}
\end{figure}

\begin{figure}[!ht]
\centering
\includegraphics[angle=270,width=9.5cm]{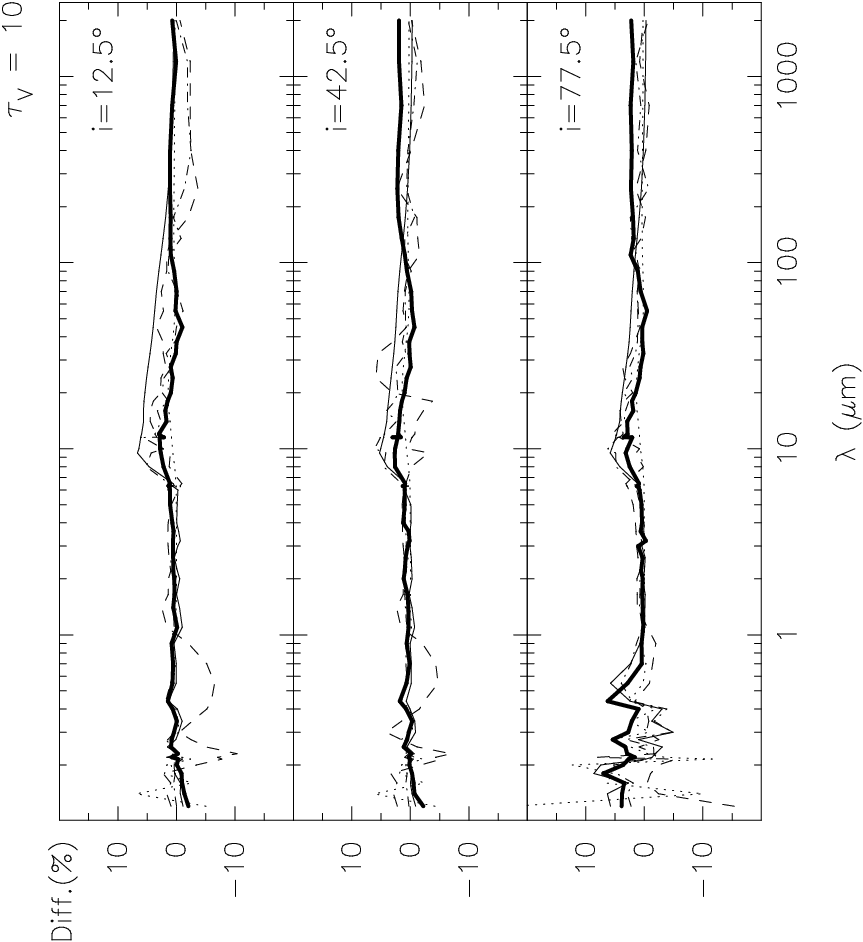}
\caption{\label{fig:diff3} Same as figure \ref{fig:diff} but for $\tau_{\nu}=10$.}
\end{figure}

\begin{figure}[!ht]
\centering
\includegraphics[angle=270,width=9.5cm]{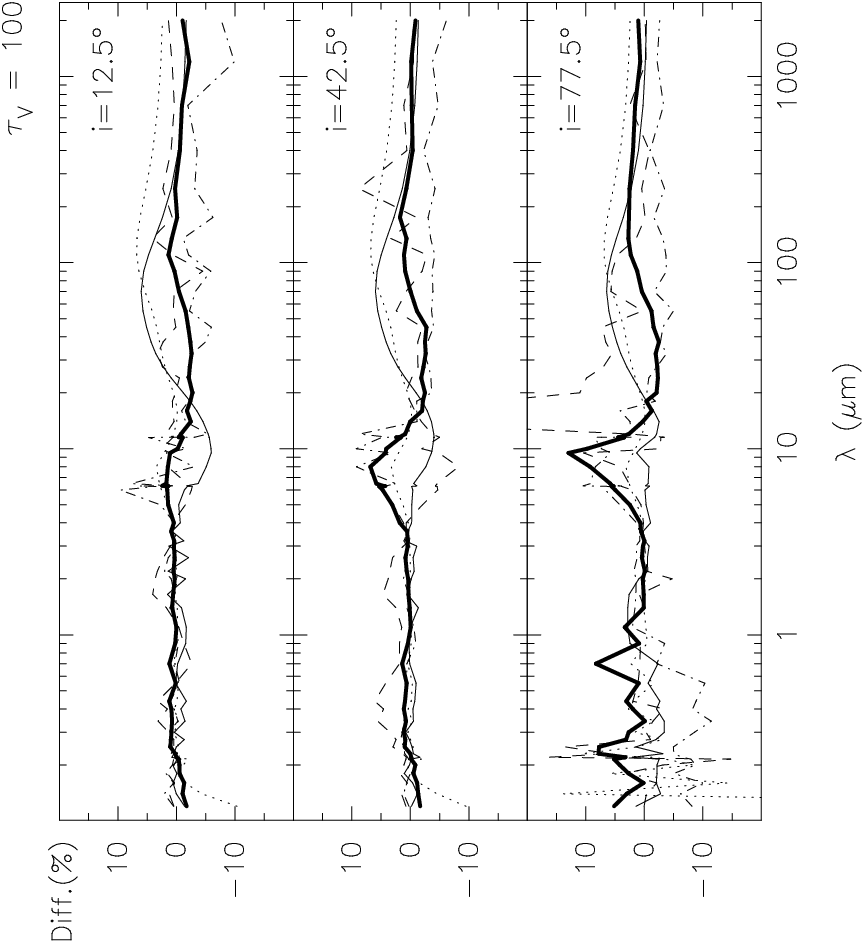}
\caption{\label{fig:diff4} Same as figure \ref{fig:diff} but for $\tau_{\nu}=100$.}
\end{figure}

\subsection{Impact of inclination angle sampling\label{sec:i_sampling}}

\begin{figure}[!ht]
\centering
\includegraphics[angle=270,width=9.5cm]{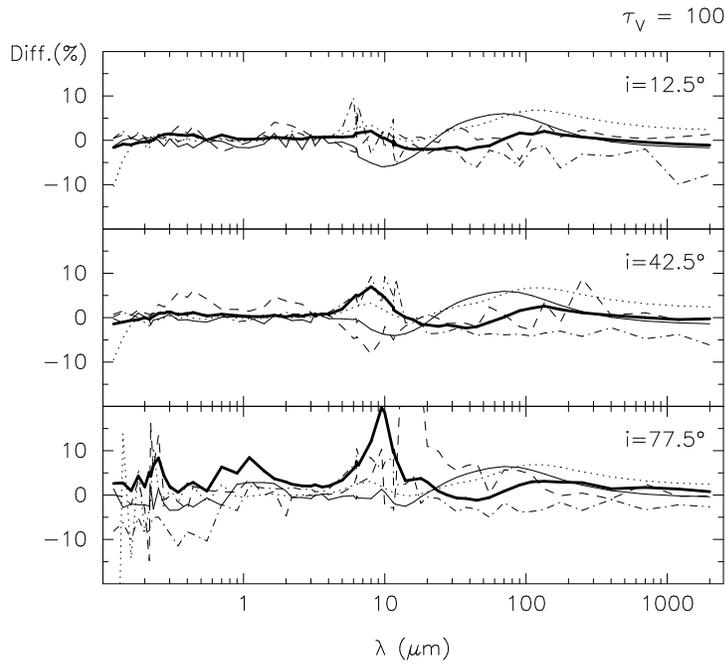}
\caption{Plots of the difference of the results MCFOST and other codes
  for the most optically thick case with high resolution on
  inclination bins. The figure presents the same calculations  as the bottom-right panel of
  Fig. \ref{fig:diff} but with 20 times narrower inclination bins. The
  lines type are the same as in Fig. \ref{fig:diff}.\label{fig:res_capt}}
\end{figure}

To verify further that an inclination range (as used) does not introduce a
noticeable bias in our comparisons, we show in Fig.~\ref{fig:res_capt}
the same calculations for inclination bins 20 times narrower. In
Fig.~\ref{fig:res_capt}, the bins run from $11.76^\circ$ to
$12.89^\circ$, $42.31^\circ$ to $42.67^\circ$and from $77.38^\circ$ to
$77.63^\circ$. The results are similar to those of
Fig.~\ref{fig:diff}, bottom right panel.

\subsection{Impact of grid sampling\label{sec:high_res}}

\begin{figure}[!p]
\centering
\includegraphics[angle=270,width=9.5cm]{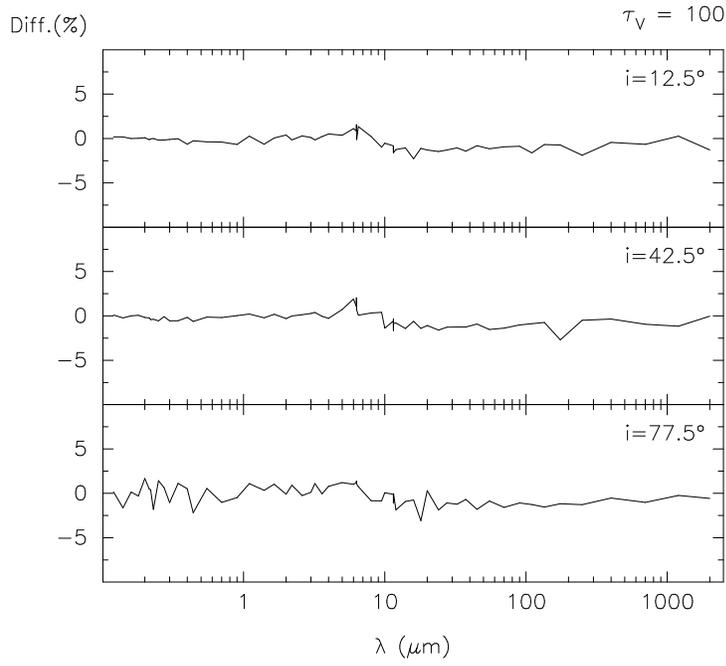}
\caption{\label{fig:sed_hr} Spatial resolution tests. Difference
  percentage between the emerging SED of the model with a grid of
  ($500 \times 200$) points (radial $\times $ vertical) and the
  reference model with a grid of ($50 \times 20$) points. The number
  of ``photons packets'' is the same for the two models. }
\end{figure}

In this section, we test the influence of the grid sampling on the
resulting SEDs.
In Fig. \ref{fig:sed_hr}, the difference between the previous
calculations with $\tau_\mathrm{V} = 100$ and the same calculations performed
with a grid sampling 10 times finer, both in radial and vertical
direction are shown. 
The number of the photons is the same for the two simulations. The
high resolution calculations were done with the same number of photons
as previously, the corresponding runtime is $\approx$ 7 hours. 
The comparison shows that the spatial resolution has very little
influence on the SED, with differences not exceeding 2.5\% even
in the worst case.

\begin{figure}[!p]
\centering
\includegraphics[angle=270,width=9.5cm]{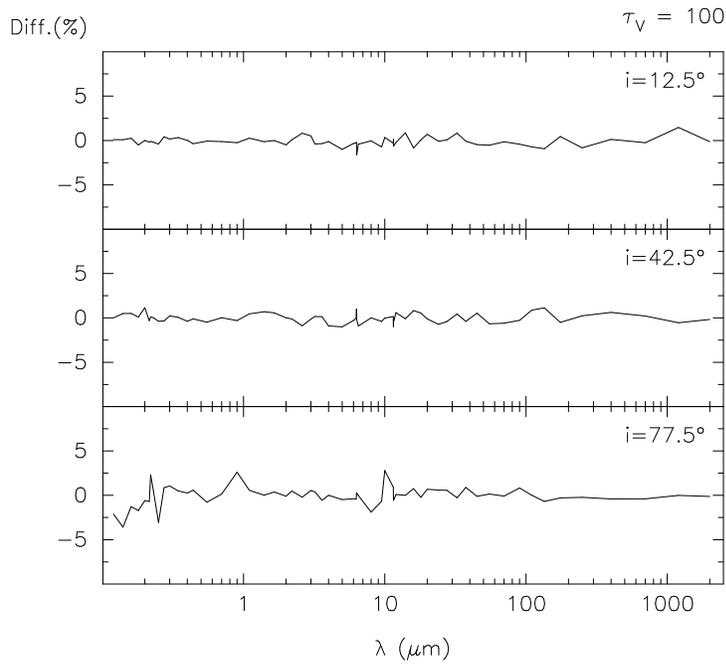}
\caption{\label{fig:sed_cutoff} Effect of the vertical cutoff. We plot the difference, given in percents, of the model with a vertical cutoff at 20 times the scale height relative to the reference model plotted in \ref{fig:sed} (cutoff at 10 times the scale height) for the most optical thick case. The two models have the same vertical resolution.}
\end{figure}

\subsection{Impact of vertical cutoff\label{sec:cutoff}}

MCFOST is the only code which uses a cylindrical grid and introduces a
vertical cutoff for the density, all other code have adopted spherical
geometries and do not need to set up a vertical cutoff.  The cutoff is
generally chosen at 10 times the local scale height of the disk. The
corresponding density is equal to $\exp(-10^2/2) \approx 2\, 10^{-22}$
times the density in the midplane of the disk and is so low that
no absorption or scattering actully occurs in this region. To
verify the influence of the cutoff on the emerging spectra, we run the
previous models with a cutoff of 20 times the scale height rather than
10, keeping the same vertical resolution and then using twice more
cells.  The difference between the models with a cutoff at 10 and 20
times the scale height are shown in Fig. \ref{fig:sed_cutoff}. They
always stay below 3 \% even at the highest optical depth.

\section{Discussion and conclusions}
In the sections above we presented a new continuum 3D radiative
transfer code, MCFOST. The efficiency and reliability of MCFOST was
tested by considering a simple and well-defined problem, a dusty disk
as described in P04, that was used as a benchmark to compare MCFOST
with 5 other RT codes. MCFOST was shown to calculate temperature
distributions and SEDs that are in excellent agreement with previous
results from the other codes. The results of MCFOST  for all the test cases are
available at the web site~:
\url{http://www-laog.obs.ujf-grenoble.fr/~cpinte/mcfost/}.

Because MCFOST reproduces well the results presented in P04, their
conclusions also hold for MCFOST. In particular, we find that the
far-infrared part of the SEDs is unaffected by the viewing angle, see
Fig. \ref{fig:sed}. On the other hand, the spectrum at shorter wavelengths is
largely modified, with the stellar spectrum dominating for pole-on
views and becoming progressively more attenuated with increasing
viewing angle. For cases with large attenuation of the starlight (high
inclinations and large optical depths) the contribution from
scattering becomes important because the dust grains have a high
albedo. Also, the spectroscopic signatures of silicates at 10$\mu$m and
20$\mu$m are visible, as expected.
The overall temperature profiles in the disk and shape of the emerging
SEDs are well reproduced by MCFOST. Differences in the radial
temperatures computed by MCFOST do not exceed 4\% with respect to the
other 5 codes in the optically thin case. For the most optically thick
case, differences do not exceed 15\%. Regarding SEDs, MCFOST is similar
to all other codes to better than 10\% for low optical
thickness. Devitations as large as 15\% are however observed in the
most defavorable case, \textit{i.e.}, high tilt and optical thickness, except
with STEINRAY which deviates from all other codes, between $10$ and
$30\mu$m (see P04 for more detailed description of these deviations). 

In this paper, disks with fairly low equatorial V-band optical depths
were calculated, $\tau_\mathrm{V}=100$. To reproduce observed edge-on disks,
MCFOST has been able so far to handle more compact, equatorially
concentrated disks with $\tau_\mathrm{V} \sim 10^6$, illustrating the
ability of MCFOST to model much more optically thick disk than the
one used for the benchmark simulation

Various tests were further performed to show the independence of the
MCFOST results with respect to grid sampling, inclination
sampling, and position of the vertical density cut-off in the disk.
Results are not noticeably affected.



\begin{thebibliography}{99}
\expandafter\ifx\csname natexlab\endcsname\relax\def\natexlab#1{#1}\fi

\bibitem[{{Bjorkman} \& {Wood}\ (2001)}]{Bjorkman01}
{Bjorkman}, J.~E. \& {Wood}, K. 2001, \apj, 554, 615

\bibitem[{{Cashwell} \& {Everett}\ (1959)}]{Cashwell}
{Cashwell}, E. \& {Everett}, C. 1959, A practical manual on the Monte Carlo
  Method for random walk problem (Pergamon, New York)

\bibitem[{{Chrysostomou} {et~al.}\ (1997)}]{Chrysostomo97}
{Chrysostomou}, A., {M\'enard}, F., {Gledhill}, T.~M., {Clark}, S., {Hough}, J. H., {McCall}, A., {Tamura}, M. 1997, \mnras,
  285, 750

\bibitem[{{Draine} \& {Lee}\ (1984)}]{Draine84}
{Draine}, B.~T. \& {Lee}, H.~M. 1984, \apj, 285, 89

\bibitem[{{Duch{\^ e}ne} {et~al.}\ (2004)}]{Duchene04}
{Duch{\^ e}ne}, G., {McCabe}, C., {Ghez}, A.~M., \& {Macintosh}, B.~A. 2004,
  \apj, 606, 969

\bibitem[{{Lucy}\ (1999)}]{Lucy99}
{Lucy}, L.~B. 1999, \aap, 345, 211

\bibitem[{{M\'enard}\ (1989)}]{Menard}
{M\'enard}, F. 1989, PhD thesis, Universit\'e de Montr\'eal

\bibitem[{{Monin} {et~al.}\ (1998)}]{Monin98}
{Monin}, J.-L., {M\'enard}, F., \& {Duch\^ene}, G. 1998, \aap, 339, 113

\bibitem[{{Pascucci} {et~al.}\ (2004)}]{Pascucci04}
{Pascucci}, I., {Wolf}, S., {Steinacker}, J., {Dullemond}, C.P.,
{Henning}, T., {Niccolini}, G., {Woitke}, P., \& {Lopez}, B. 2004, \aap, 417, 793

\bibitem[{{Pinte} \& {M{\' e}nard}\ (2004)}]{Pinte04}
{Pinte}, C. \& {M{\' e}nard}, F. 2004, in AIP Conf. Proc. 713: The Search for
  Other Worlds, 123--126

\bibitem[{{Solc}\ (1989)}]{Solc89}
{Solc}, M. 1989, Astronomische Nachrichten, 310, 329

\end{thebibliography}
\end{document}